\documentclass[12pt]{iopart}
\usepackage{graphicx}
\usepackage{ulem}
\usepackage{color}
\usepackage{iopams}
\usepackage{setstack}
\usepackage{mathabx}

\begin{document}

\title{Bayesian Mechanics of Synaptic Learning under the Free Energy Principle}
\author{Chang Sub Kim}
\address{Department of Physics,
Chonnam National University,
Gwangju 61186, Republic of Korea}
\ead{cskim@jnu.ac.kr}

\begin{abstract}
The brain is a biological system comprising nerve cells and orchestrates its embodied agent's perception, behavior, and learning in the dynamic environment.
The free energy principle (FEP) advocated by Karl Friston explicates the local, recurrent, and self-supervised neurodynamics of the brain's higher-order functions.
In this paper, we continue to finesse the FEP through the physics-guided formulation; specifically, we apply our theory to synaptic learning by considering it an inference problem under the FEP and derive the governing equations, called Bayesian mechanics.
Our study uncovers how the brain infers weight change and postsynaptic activity, conditioned on the presynaptic input, by deploying the generative models of the likelihood and prior belief.
Consequently, we exemplify the synaptic plasticity in the brain with a simple model: we illustrate that the brain organizes an optimal trajectory in neural phase space during synaptic learning in continuous time, which variationally minimizes synaptic surprisal.
\end{abstract}

\vskip1in
\maketitle

\section{Introduction}
\label{Introduction}
The brain's perception, body movement, and learning are conjointly organized to ensure the embodied agents' homeostasis and adaptive fitness in the environment.
It is tempting to imagine a neural observer in the brain presiding over higher animals' cognitive control.
Such a homunculus idea is untenable and must be discarded in the present-day brain theory \cite{Crick:2003}.
However, there is still a much distance to the complete scientific understanding of the emergent higher-order functions from the brain matter; it demands a comprehension of the profound interplay between the scientific reductionism and teleological holism standpoints \cite{Clark:2013,Buzsaki:2019}.

The brain-inspired FEP is a purposive theory that bridges the gap between top-down teleology and bottom-up scientific constructionism.
According to the FEP \cite{Friston:2010,Friston:2023}, all living systems are self-organized to tend to avoid an atypical niche in the environment for existence.
The FEP adopts the autopoietic hypothesis \cite{Maturana:1980} and scientifically formalizes the abductive rationale of organisms' making optimal predictions and behavior from incomplete sensory data.
To be precise, the FEP suggests an information-theoretic, variational measure of environmental atypicality, termed \textit{free energy} (FE).
The FE objective is technically defined as a functional of the probabilistic generative density specifying the brain's internal model of sensory-data generation and environmental change and an online auxiliary density actuating variation.
The Bayesian brain computes the posterior of the environmental causes of uncertain sensory data by minimizing the FE, whose detailed continuous-state description can be found in \cite{Kim:2017}.
For discrete-state models of the FEP with discrete time, we recommend \cite{DaCosta:2020,Smith:2022} to readers.

When a Gaussian probability is employed for the variational density \cite{Friston:2007a}, the FE becomes a $L_2$ norm specified by the Gaussian means and variances and termed the Laplace-encoded FE \cite{Kim:2017}.
Thus, the Laplace-encoded FE provides a scientific base of the $L_2$ objectives in a principled manner, which are widely used in machine learning and artificial intelligence.
For instance, the optimization function in the predictive-coding framework is proposed to be a sum of the squared prediction errors \cite{Rao:1999}.
Also, the loss function of a typical artificial neural network (ANN) is often written as a sum of squared differences between the ground truth and the predictive entries from the network \cite{LeCun:2015}.
Furthermore, it is argued that the Gaussian sufficient statistics are encoded by the biophysical brain variables, which form the brain's low-dimensional representations of environmental states.
This way the brain acquires access to the encoded FE for minimization as it becomes fully specified in terms of the brain's internal states.

Our research over the years has been devoted to developing continuous-state implementation of the FE minimization in a manner guided by physics laws and principles \cite{Kim:2018,Kim:2021,Kim:2023}.
We endeavored to advance the FEP to the point where it coalesces into a unified principle of top-down architecture and material base.
Moreover, to promote the FEP to nonstationary problems, we incorporated the fact that the physical brain is in a nonequilibrium (NEQ) stationary state and is generally continually aroused by nonstationary sensory stimuli.
The functional brain must perform the variational Bayesian inversion of nonstationary sensory data to compute the posterior mentioned above.
Previously, we accounted for the brain behavior of perception and motor control as described by attractor dynamics and termed the governing equations \textit{Bayesian mechanics} (BM).
The BM coordinates the brain's sensory estimation and motor prediction in neural phase space.
In this paper, we make further progress by incorporating the brain's \textit{synaptic learning} into the BM, which we did not accommodate in our earlier studies.
Learning constitutes the crucial brain function of consolidating memory, e.g., via Hebbian plasticity \cite{Hebb:1949}.

This paper aims to provide a simple but insightful model for synaptic learning in the brain.
Our agendas are that the functional brain operates continually using continuous environmental representations and that synaptic learning is a cognitive phenomenon that may very well be understood when statistical-physical laws guide it.
The notion \textit{cognition} throughout this paper is meant to be the brain's higher-order capability that involves a top-down, internal model.
We consider the NEQ brain a problem-solving matter, cognitively interacting with the environment.
To quantify the synaptic cognition, we will specify the generative densities furnishing the Laplace-encoded FE in a manner to meet the NEQ stationarity and present the FE minimization scheme by practicing the principle of least action (Hamilton's principle) \cite{Landau:1976}.
The novel contributions worked out in this paper are discussed in Section~\ref{Discussion}.

The rest of the paper is organized as follows.
In Section~\ref{Single_Synapse}, the single-synapse structure of our interest is described.
The essence of the FEP is recapitulated with revision made for synaptic learning in Section~\ref{FEP_Recap}.
In Section~\ref{NEQ_Densities}, an NEQ formulation is presented, which determines the likelihood and prior densities in the physical brain.
Next, Section~\ref{BM_Derivation} identifies the FE objective as a classical action and derives the governing equations of synaptic dynamics by exercising the Hamilton principle.
The utility of our theory is demonstrated in Section~\ref{Application} using a simple model.
After the discussion in Section~\ref{Discussion}, a conclusion is given in Section~\ref{Conclusion}.

\section{Single Synapse Model}
\label{Single_Synapse}
This work concerns the brain's synaptic learning without considering how environmental processes arouse stimuli at the sensory interface; as a parsimonious model, we focus on a single synapse within the brain's internal environment.
For instance, in the hippocampus, the postsynaptic action potential in the dentate gyrus is evoked by a presynaptic signal from the entorhinal cortex caused by a neural signal from other brain areas.
Accordingly, the synaptic coupling between two pyramidal neurons in the hippocampus constitutes a single synaptic assembly of interest.

\begin{figure}[t]
\begin{center}
\includegraphics[width=10cm]{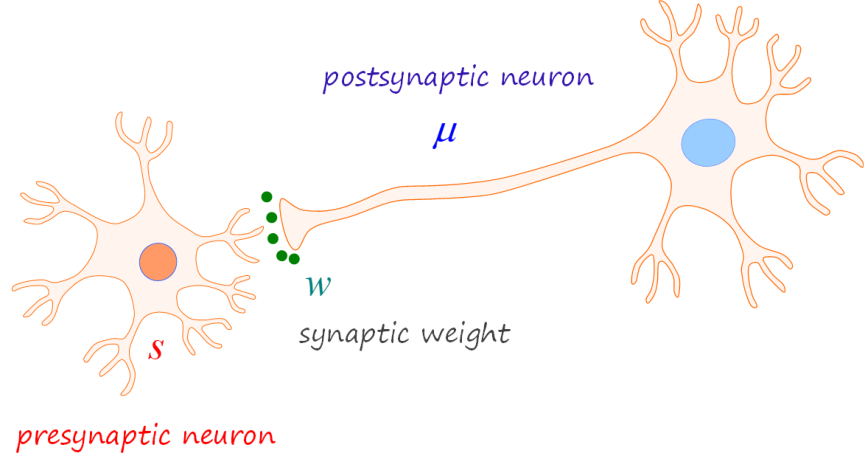}
\caption{Single synaptic assembly. The postsynaptic neural state $\mu$ is neurophysically evoked by the presynaptic signal $s$, mediated by the weight change $\Delta w$ according to Hebb's rule $\propto s\mu$. We adopt the Bayesian-inference perspective, suggesting that the brain state $\mu$ infers the cause of the presynaptic input $s$, and the weight state $w$ makes up the synaptic input-output interface.
\label{Synapse}}
\end{center}
\end{figure}

We depict the single synaptic model in Figure~\ref{Synapse}, where the presynaptic and postsynaptic signals are denoted by $s$ and $\mu$, respectively; both are the brain's representations of noisy synaptic signals.
In addition, the synaptic plasticity is mediated by the weight strength denoted by $w$.
The synaptic structure considered is generic for all neurons; accordingly, the ensuing formulation below applies to other brain regions.
Note that we will handle the weight variable $w$ as a neurophysical degree of freedom like $s$ and $\mu$; this handling contrasts with ANN models, where the weights are treated as a static parameter.

\section{Free Energy Principle for Synaptic Learning}
\label{FEP_Recap}
The brain-inspired FEP is built on three hypotheses: 1) surprisal hypothesis, 2) representation hypothesis, and 3) computability hypothesis, which we recapitulate here with the revision applying to the synaptic learning problem.

\subsection{Surprisal Hypothesis}
\label{Hypothesis1}
We assume that the presynaptic signals $s$ streaming into the synaptic interface are prescribed and focus on the resulting synaptic dynamics.
For convenience, here we introduce the notation $\tilde\vartheta$, by which we collectively denote the postsynaptic variable ${\cal M}$ and the weight variable ${\cal W}$:
\[ \tilde\vartheta = \{{\cal M},{\cal W}\}.\]
The variables ${\cal M}$ and ${\cal W}$ are stochastic and unknown, so they are hidden from the brain's perspective.

The brain's cognitive goal is to compute the posterior $p(\tilde\vartheta|s)$, which the FEP fulfills via variational Bayes in the following manner:
First, we define the information-theoretic measure called the Kullback-Leibler (KL) divergence:
\begin{equation}
\label{KL}
D_{KL}\left(q(\tilde\vartheta)\|p(\tilde\vartheta|s)\right) = \int d\tilde\vartheta q(\tilde\vartheta)\ln\frac{q(\tilde\vartheta)}{p(\tilde\vartheta|s)},
\end{equation}
which is always positive \cite{Cover:1991}, where $d\tilde\vartheta$ means $d{\cal M}d{\cal W}$.
\textit{Some terminologies}: $q(\tilde\vartheta)$ is called R-density, which approximates the true posterior $p(\tilde\vartheta|s)$ in the variational scheme.
The posterior makes up the so-called G-density $p(\tilde\vartheta,s)=p(\tilde\vartheta|s)p(s)$ together with the marginal density $p(s)$ \cite{Kim:2017}.
Second, using the preceding product rule, the above KL divergence can be decomposed to
\[
D_{KL}\left(q(\tilde\vartheta)\|p(\tilde\vartheta|s)\right) =  {\cal F}[q(\tilde\vartheta),p(\tilde\vartheta,s)] + \ln p(s).
\]
The functional $\cal F$ on the right-hand side (RHS), which is identified to be
\begin{equation}
\label{IFE}
{\cal F}[q(\tilde\vartheta),p(\tilde\vartheta,s)] \equiv \int d\tilde\vartheta~ q(\tilde\vartheta)\ln\frac{q(\vartheta)}{p(\tilde\vartheta,s)},
\end{equation}
is the \textit{informational} free energy (FE).
Third, the positivity of $D_{KL}$ leads to the inequality
\begin{equation}
\label{FEP1}
-\ln p(s) \le {\cal F}[q(\tilde\vartheta),p(\tilde\vartheta,s)].
\end{equation}

Equation~(\ref{FEP1}) is the mathematical statement of the brain-inspired FEP accounting for life and cognitive phenomena in a universal manner, which comprises the \textit{surprisal hypothesis}.
In the present context, the preceding inequality enunciates that synaptic learning corresponds to the brain's minimizing ${\cal F}$, which is a proxy for \textit{synaptic surprisal}, $-\ln p(s)$, as an upper bound.
In practice, it is intractable to determine the marginal density $p(s)$, which provides \textit{synaptic evidence} to the brain.
Note here that ${\cal F}$ is called FE by mimicking thermodynamic FE in physics, which monotonically decreases upon spontaneous changes in a macroscopic open system, conforming to the second law of thermodynamics \cite{Kim:2023}.

\subsection{Representation Hypothesis}
\label{Hypothesis2}
According to the inequality [Equation~(\ref{FEP1})], the brain variationally minimizes ${\cal F}$ by means of the R-density $q(\tilde\vartheta)$:
when the synaptic interface is elicited by the presynaptic stream $s$, the brain launches $q(\tilde\vartheta)$, an online approximation of the posterior;
in the face of the synaptic stream, the R-density probabilistically represents the uncertain, hidden causes $\tilde\vartheta=\{{\cal M},{\cal W}\}$, trying to match best with the posterior.
Here, we adopt the Laplace approximation for the R-density, which assumes a Gaussian form \cite{Tierney:1986}:
\begin{equation}\label{Laplace}
q(\tilde\vartheta) = \frac{1}{\sqrt{2\pi\tilde\sigma^2}}\exp[-\frac{1}{2\tilde\sigma^2}(\tilde\vartheta-\tilde\mu)^2],
\end{equation}
where $\tilde\mu$ and $\tilde\sigma$ are the sufficient statistics of the Gaussian density.
In particular, we intend that the means denoted by
\[\tilde\mu = \{\mu,w\}\]
are the coarse-grained representations of high-dimensional $\tilde\vartheta=\{{\cal M},{\cal W}\}$;
they are the latent brain variables in low dimensional neural space \cite{Chaudhuri:2019}.
Also, the dependence on the variances $\tilde\sigma$ can be eliminated by further manipulation as elaborated in \cite{Kim:2017}.
Then, under the Laplace approximation, the FE functional reduces to
\[ {\cal F}[q(\tilde\vartheta),p(\tilde\vartheta,s)] = -\ln p(\tilde\mu,s) + {\rm constants}. \]
The nontrivial part in the reduced expression is the Laplace-encoded FE denoted by $F$:
\begin{equation}
\label{Laplace-FE}
F(\mu,w,s) = -\ln p(\mu,w,s),
\end{equation}
which is a function of only the brain variables $\mu$ and $w$.
In the present work, the presynaptic input $s$ is not a dynamical variable but is handled as an external time-dependent input.

Here, the brain is assumed to be endowed with the generative density $p(\mu,w,s)$ encoded over the evolutionary and developmental time scales.
The FEP inequality given in Equation~(\ref{FEP1}) now becomes
\begin{equation}
\label{FEP2}
-\ln p(s) \le F(\mu,w,s);
\end{equation}
the brain has an access to $F(\mu,w,s)$ by means of its internal variables $\mu$ and $w$.
The preceding expression comprises the \textit{representation hypothesis} in the FEP; the brain uses the coarse-grained representations $\mu$ and $w$ in variationally minimizing synaptic surprisal.
Then, by applying the product rule $p(\mu,w,s) = p(w|\mu,s)p(\mu,s)$, the Laplace-encoded FE is completed as
\begin{equation}\label{LFE}
F(\mu,w|s) = - \ln p(w|\mu,s)p(\mu,s),
\end{equation}
where $p(w|\mu,s)$ is the likelihood of the weight strength $w$ given a postsynaptic signal $\mu$, and $p(\mu,s)$ is the prior about the postsynaptic dynamics, both subject to the presynaptic input $s$.

Equation~(\ref{LFE}) is the objective function for synaptic learning under the  FEP, furnished with only brain variables, which makes the brain-inspired FEP a biologically plausible theory.
Previously, we suggested that all the involved probabilities be specified as NEQ stationary densities derived from the Fokker--Planck equation \cite{Kim:2023}.
This work takes a different approach to determining the NEQ densities in Section~\ref{NEQ_Densities}.

\subsection{Computability Hypothesis}
\label{Hypothesis3}
The brain is endowed with the mechanism that actuates the FE minimization, which comprises the \textit{computability hypothesis} in the FEP.
The conventional continuous-state implementation assumes that the brain employs gradient descent (GD) methods to execute the FE minimization \cite{Friston:2023}.
The GD schemes update the neural activity $\mu$ downhill on the FE landscape, which is woven by the generalized coordinates of motion of all dynamical orders surpassing the second order, namely, acceleration \cite{Friston:2007b,Friston:2008}.
It is argued that generalized motion can effectively incorporate the temporal correlation of random fluctuations in stochastic dynamics beyond white noise.
However, the idea of generalized motion transcends normative Newtonian physics; thus, its theoretical ground draws critical attention in the literature \cite{Aguilera:2021,Kim:2021}.
For the weight variable $w$, to incorporate its slower change than the neural activity, a different update rule is applied: for instance, instead of the weight (parameters or hyper-parameters), its rate may be updated under the GD scheme \cite{Kim:2017}.
Recently, researchers have extended the applicability of FEP-based GD algorithms to robotics and artificial intelligence problems, emphasizing colored-noise modeling \cite{Meera:2021}.
However, it is significant to note that using GD methods is not legitimate when the environmental inputs vary fast so that the FE landscape becomes non-static (see, for further discussion, Section~\ref{Discussion}).
Our formulation aims at the general time-dependent situation and, thus, avoids using a GD scheme; instead, we identify the FE objective to be a classical action in mechanics and exercise Hamilton's principle for the FE minimization according to the standard theory \cite{Landau:1976}.
The details are given in Section~\ref{BM_Derivation}, where we derive the governing equations of motion for synaptic inference regarding the canonical physical variables without invoking the generalized motion.

\section{Nonequilibrium Generative Densities}
\label{NEQ_Densities}
We argued that the physical brain probabilistically encodes the representations of the internal and external hidden states (Section~\ref{Hypothesis2}).
The encoded probabilities constitute the generative densities that furnish the brain with FE objective for variational Bayesian inference.
Therefore, the generative densities must be specified in terms of the biophysical brain variables in an NEQ stationary state.
Here, we present a stochastic thermodynamic model for the NEQ densities, viewing the brain as a soft material consisting of neural constituents.
This perspective brings us closer to understanding the brain's NEQ states.

\subsection{Prior for Postsynaptic Activity}
\label{NEQ_Prior}
For a simple description, we assume that the brain variable $\mu$ obeys an overdamped Langevin dynamics on a mesoscopic scale:
\begin{equation} \label{State-eq}
\frac{d\mu}{dt} = f(\mu,\theta) + \xi,
\end{equation}
where $f$ and $\xi$ on the RHS are the deterministic and random forces, respectively, causing the neural change; $\theta$ encapsulated in $f$ denotes an input parameter affecting the system dynamics.
The solution to Equation~(\ref{State-eq}) describes a stochastic path or trajectory $\mu=\mu(t)$ in continuous state space.
Recall that the neural variable $\mu$ is the mean of the R-density probabilistically representing external environmental states online, which may be viewed as a mean field.
Also, it is evident that the state transition between two arbitrarily-close times described by Equation~(\ref{State-eq}) is Markovian.
Further assumptions imposed are i) the noise $\xi$ is Gaussian about zero mean, rendering $\langle\xi\rangle=0$, and ii) the noise is delta-correlated, a.k.a. white, through
\begin{equation}\label{white}
\langle\xi(t^\prime)\xi(t)\rangle = \sigma_\mu^2 \delta(t^\prime-t),
\end{equation}
where $\sigma_\mu^2$ is the noise strength.
Strictly considering, the biological brain is in an NEQ stationary state, whose temperature $T$ is distinct from the environmental value; however, here, we consider that the brain is locally in equilibrium characterized by its body temperature.
Also, we assume that the noise strength is given, according to the fluctuation-dissipation theorem \cite{Kubo:1991}, as
\begin{equation}\label{noise-strength}
\sigma_\mu^2 = 2\gamma_\mu^{-1} k_BT,
\end{equation}
where $\gamma_\mu$ is the frictional coefficient of the brain matter, and $k_B$ is the Boltzmann constant.

Under the prescribed assumptions, we build the transition probability along a trajectory $\mu=\mu(t)$ as time $t$ elapses.
To proceed with the derivation, we first note a technical subtlety involved in the white noise $\xi(t)$: it is mathematically ill-defined because the variance is divergent [see Equation~(\ref{white})].
To address this, the Wiener process, defined through $\Delta W\equiv\xi \Delta t$, is often conceived.
This process introduces a form of coarse-graining over a short time interval $\Delta t$, effectively bypassing the singularity of the white noise at an instant time.
The Wiener process is also Gaussian about zero mean with the well-defined variance $\langle (\Delta W)^2\rangle = \sigma_\mu^2 \Delta t$.
However, one must pay the price for the Wiener recipe when the Riemann integral is performed for state functions over a finite-time elapse.
In our derivation, we adopt the Ito convention that interprets the integral of Equation~(\ref{State-eq}) over the time interval $\Delta t=t_{n+1}-t_n$ as
\[\Delta\mu_n = f(\mu_n,\theta_n)\Delta t + \Delta W_n,\]
where the first term on the RHS was approximated by choosing the value for $f(\mu)$ at the initial time $t_n$; other terms are $\Delta\mu_n=\mu_{n+1}-\mu_n$ and $\Delta W_n=W_{n+1}-W_n$.
Next, using the Gussianity of $\Delta W_n$, we define the transition probability $p(n+1|n)$ from the Wiener state $W_n$ to the next $W_{n+1}$ as \cite{Adib:2008}
\begin{equation*}
p(n+1|n)\simeq \exp\left\{-\frac{1}{2\sigma_\mu^2 \Delta t}\Big(\Delta\mu_n-f(\mu_n,\theta_n) \Delta t\Big)^2 \right\}.
\end{equation*}
Then, the full Markovian transition over $N(=t/\Delta t)$ time steps during the finite time $0\le t^\prime\le t$ can be built as
\[
\prod_{n=0}^{N-1} p(n+1|n)\simeq \exp\left\{-\frac{\Delta t}{2\sigma_\mu^2}\sum_n\Big(\frac{\Delta\mu_n}{\Delta t}-f(\mu_n,\theta_n)\Big)^2 \right\}.
\]
As a final step, we take the continuous limit $\Delta t\rightarrow 0$ in the preceding expression and obtain the path probability $p(\mu,\theta)$, up to a normalization constant, as
\begin{equation}\label{OMmu}
p(\mu,\theta) \sim \exp\left\{ -\frac{1}{2\sigma_\mu^2}\int_0^t dt^\prime\left(\frac{d\mu}{dt^\prime} -f(\mu,\theta(t^\prime))\right)^2 \right\},
\end{equation}
which is known as the Onsager-Machlup function \cite{Hunt:1981}.

The above Onsager--Machlup expression specifies the transition probability of the neural state $\mu$, given initial condition $\mu(0)$, along the continuous path $\mu=\mu(t)$.
When the parameter $\theta$ is replaced with $s$, it represents the prior density $p(\mu,s)$ in Equation~(\ref{LFE}) accounting for the brain's belief about or already-acquired knowledge of how the postsynaptic activity $\mu$ behaves.

\subsection{Likelihood of Synaptic Change}
\label{NEQ_Likelihood}
Neurotransmitter transport at the synaptic interface mediates synaptic coupling between two neurons, which is often effectively described by the weight variable $w$.
We assume that the brain is endowed with an internal model of weight dynamics leveraging learning; learning constitutes the crucial brain function of consolidating memory, e.g., via long-term potentiation.

We consider the synaptic weight $w$ a time-dependent variable rather than a static parameter, and the synaptic plasticity is described by its the rate $\dot w=dw/dt$.
We propose that similar to Equation~(\ref{State-eq}), the synaptic plasticity is governed by the stochastic equation:
\begin{equation}
\label{Weight-eq}
\frac{dw}{dt} = h(w,\theta) + \chi,
\end{equation}
where $h$ is the biophysical force causing the weight change, and $\chi$ is the additive white noise associated with the synaptic process; again, $\theta$ is a time-dependent input parameter.
The noise is assumed to be Gaussian about zero mean and delta-correlated: $\langle\chi(t)\chi(t^\prime)\rangle = \sigma_w^2 \delta(t-t^\prime)$ with $\sigma_w^2$ being the noise strength.

Next, to smooth the temporal singularity associated with the white noise $\chi$, we consider the Wiener process $\Delta W=\chi\Delta t$, which is also Gaussian about zero mean with the well-defined variance, $\langle (\Delta W)^2\rangle = \sigma_w^2 \Delta t$.
Then, we proceed with the same formulation with Section~\ref{NEQ_Prior} to specify the NEQ likelihood density $p(w|\theta)$.
The result is given as
\begin{equation}\label{OMw}
p(w|\theta) \sim \exp\left\{ -\frac{1}{2\sigma_w^2}\int_0^t dt^\prime\left(\frac{dw}{dt^\prime} -h(w,\theta(t^\prime))\right)^2 \right\},
\end{equation}
which represents the Onsager--Machlup transition probability along the continuous path $w=w(t)$, subject to initial condition $w(0)$.

In obtaining the above likelihood and prior densities, Equations~(\ref{OMmu}) and (\ref{OMw}), respectively, we assumed that the random fluctuations in the neuronal dynamics were delta-correlated, i.e., white noises.
The brain signals, by contrast, evidently reveal the frequency spectrum reflecting color-correlated dynamics \cite{Buzsaki:2004}, which supports the criticality idea in the brain \cite{Beggs:2008}.
In this work, we consider only the ideal white noise for a practical illustration of determining the NEQ brain densities in a physics-grounded manner.
To obtain an analytic expression for the NEQ densities is intractable under general conditions even in the steady state \cite{Kim:2023}; they are usually assumed to be an instant Gaussian set by the Gaussian random noises imposed on the Langevin description \cite{Friston:2010,Friston:2023}.

\section{Bayesian Mechanics: Computability of Synaptic Learning}
\label{BM_Derivation}

The FE landscape becomes nonstatic when the input parameter $\theta$ in the generative densities [Equations~(\ref{OMmu}) and (\ref{OMw})] is explicitly time-dependent.
In this case, it is anticipated that the GD implementation on the FE landscape will fail.
Here, we formulate the brain's computability under nonstationary conditions, facilitating nonautonomous neural computation.

In the synaptic learning problem, the presynaptic signal $s$ acts as the input parameter $\theta$.
Accordingly, we replace $\theta$ with $s$ in the Onsager--Maclup representations for the NEQ densities and substitute the results into Equation~(\ref{LFE}) to obtain the Laplace-encoded FE.
The outcome is given as
\begin{equation}\label{wAction}
F = \int_0^t {\cal L}(\mu,w;\dot \mu,\dot w;s) dt^\prime,
\end{equation}
where the integrand ${\cal L}$ is expressed as
\begin{equation}
\label{learning_Lag}
{\cal L}(\mu,w;\dot \mu,\dot w;s) \equiv \frac{1}{2\sigma_\mu^2}\left(\frac{d\mu}{dt^\prime} -f(\mu,w;s(t^\prime))\right)^2 + \frac{1}{2\sigma_w^2}\left(\frac{dw}{dt^\prime} -h(\mu,w,s(t^\prime))\right)^2.
\end{equation}
Note that in Equation~(\ref{learning_Lag}), we concretely displayed the autonomous dependence on the variables $\mu$ and $w$ and the nonautonomous dependence on the input $s$ through the generative functions $f$ and $h$.

Equation~(\ref{wAction}) manifests a specific association of the FE objective $F$ with the mathematical object ${\cal L}$; namely, $F$ is given as a time integral of ${\cal L}$.
This observation is reminiscent of the relation between the action and Lagrangian in classical mechanics \cite{Landau:1976}.
Accordingly, by analogy, if we identify $F$ as an effective \textit{action} ${\cal S}$ and the integrand ${\cal L}$ as an effective \textit{Lagrangian} for the brain's cognitive computation, the FE minimization, which is mathematically performed by $\delta F = 0$ under the FEP, is precisely mapped to exercising Hamilton's principle, $\delta {\cal S} = 0$.
Then, the Euler-Lagrange equations of motion for determining the optimal trajectories $\mu(t)$ and $w(t)$ will follow straightforwardly, constituting the synaptic BM.
Note the temperature dependence of the Lagrangian [Equation~(\ref{learning_Lag})] via the noisy strengths $\sigma_\mu^2$ and $\sigma_w^2$, see Equation~(\ref{noise-strength}), which makes ${\cal L}$ a \textit{thermal} Lagrangian \cite{Hunt:1981}.

Here, working in the Hamiltonian description is more suitable for our purposes.
To this end, we carried out a Legendre transformation to derive an effective \textit{Hamiltonian} ${\cal H}$; the outcome is expressed as
\begin{equation}
\label{learning_Ham}
{\cal H} = \frac{p_\mu^2}{2m_\mu} + \frac{p_w^2}{2m_w} + p_\mu f(\mu,w;s) + p_w h(w,\mu;s).
\end{equation}
In the preceding expression of Hamiltonian, the new variables $p_\mu$ and $p_w$ appear, which are mechanically conjugate to the variables $\mu$ and $w$, respectively; they are determined from the definitions:
\begin{equation}\label{momenta}
p_\mu = \frac{\partial{\cal L}}{\partial \dot\mu} \quad{\rm and}\quad  p_w = \frac{\partial{\cal L}}{\partial \dot w}.
\end{equation}
Also, the constants, $m_\mu$ and $m_w$ were defined to be
\begin{equation}
\label{masses}
m_\mu= 1/\sigma_\mu^2 \quad{\rm and}\quad m_w= 1/\sigma_w^2,
\end{equation}
which are a measure of respective precision of the probabilistic generative models, Equations~(\ref{OMmu}) and (\ref{OMw}).
Equation~(\ref{noise-strength}) suggests that the generative precisions are a biophysical constant specified by the body temperature and the friction of the brain matter.
A few points about the Hamiltonian ${\cal H}$ are noteworthy:
The variables ($\mu$, $w$) and ($p_\mu$, $p_w$) correspond to \textit{positions} and \textit{momenta}, respectively, and the generative precisions $m_\mu$ and $m_w$ may be interpreted as a \textit{neural mass} as a metaphor.
The Hamiltonian is not breakable into the kinetic and potential energies because the third and fourth terms on the RHS in Equation~(\ref{learning_Ham}) are given as a product of momentum and position variables.
The ${\cal H}$ function does not furnish a conservative-energy surface because of its explicit time dependence through the presynaptic signal $s(t)$, which makes synaptic learning a nonautonomous problem.

The generative functions $f$ and $h$ for synaptic learning were introduced in Equations~(\ref{State-eq}) and (\ref{Weight-eq}) without specifying them; they are the biophysical forces driving synaptic dynamics at the neuronal level.
We now specify them by the following models:
\begin{eqnarray}
f(\mu,w;s) & = & -\gamma_\mu(\mu-\mu_d) + ws,\label{learning_NGmu}\\
h(\mu,w;s) & = & -\gamma_w(w-w_d) + s\mu.\label{learning_NGw}
\end{eqnarray}
The first terms on the RHSs, involving the damping coefficients $\gamma_\mu$ and $\gamma_w$, prevent an unlimited growth of $\mu$ and $w$ \cite{Miller:1994}.
The linear damping models may be replaced with a nonlinear alternative; for instance, the modified $- \gamma_w s^2(w-w_d)$ may be used in Equation~(\ref{learning_NGw}) \cite{Oja:1982}.
The second term $ws$ on the RHS of Equation~(\ref{learning_NGmu}) describes the presynaptic evoking weighted by $w$.
Moreover, the term $s\mu$ in Equation~(\ref{learning_NGw}) accounts for Hebb's rule; one can explore anti-Hebbian learning by inverting its sign.
The extra parameters $\mu_d$ and $w_d$ are the steady-state values of $\mu$ and $w$, respectively, without driving terms $ws$ and $s\mu$.
After substituting Equations~(\ref{learning_NGmu}) and (\ref{learning_NGw}) into Equation~(\ref{learning_Lag}) and by evaluating Equation~(\ref{momenta}), one can determine the neural representations of the momenta $p_\mu$ and $p_w$.
The results are given as
\begin{eqnarray}
& p_\mu = m_\mu(\dot\mu - f), \label{prerrormu}\\
& p_w = m_w(\dot w - h). \label{prerrorw}
\end{eqnarray}
Note that momentum represents the discrepancy between the state rate and its prediction from the generative model, modulated by precision, which corresponds to \textit{prediction error} in predictive coding theory (see discussion in Section~\ref{Discussion}).

Having specified the synaptic Hamiltonian given in Equation~(\ref{learning_Ham}), we now derive Hamilton's equations of motion by practicing the standard procedure \cite{Landau:1976}.
Here, we present only the outcome without intermediate steps:
\begin{eqnarray}
\dot \mu & =& \frac{1}{m_\mu}p_\mu -\gamma_\mu(\mu-\mu_d) + ws,\label{wHam1}\\
\dot w & =& \frac{1}{m_w}p_w -\gamma_w(w-w_d) + s \mu,\label{wHam2}\\
\dot p_\mu & =& \gamma_\mu p_\mu - s p_w,\label{wHam3} \\
\dot p_w & =& \gamma_w p_w - s p_\mu.\label{wHam4}
\end{eqnarray}
The resulting Equations~(\ref{wHam1})--({\ref{wHam4}) are a set of coupled differential equations for four dynamical variables $\mu$, $w$, $p_\mu$, and $p_w$, subject to the time-dependent input source $s$, which constitute the synaptic BM governing \textit{co-evolution} of the state and weight variables.
In Figure~\ref{Circuitry}, we show the neural circuitry implied by the derived BM.
We argue that the functional behavior depicted in the circuitry is generic in every synapse in the brain like every cortical column in the neocortex behaves as a sensorimotor system performing the same intrinsic function \cite{Hawkins:2017}.

\begin{figure}[t]
\begin{center}
\includegraphics[width=7cm]{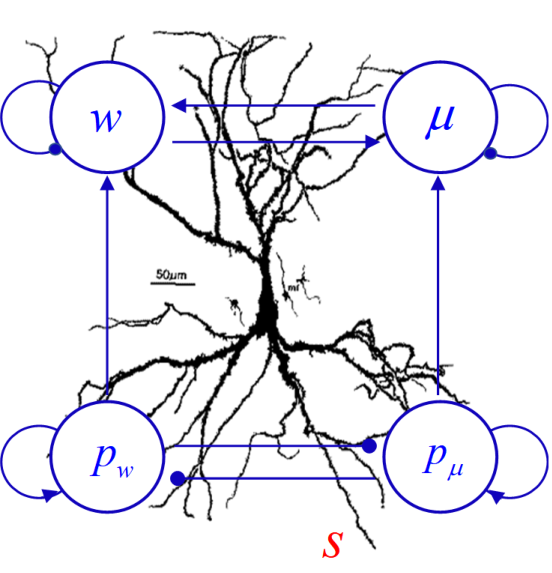}
\caption{
Schematic of the neural circuitry. The diagram manifests the workings of the synaptic BM: the presynaptic input $s(t)$ drives the interconnected, recurrent dynamics among the state $(w,\mu)$ and momentum $(p_w,p_\mu)$ variables. The links depicted by arrowheads indicate an excitatory coupling within a neural unit or between two neural units, whereas the dot-head links indicate an inhibitory coupling.
\label{Circuitry}}
\end{center}
\end{figure}

For a more compact description, we shall define the cognitive state $\Psi$ as a column vector in four-dimensional \textit{phase space}:
\[\Psi^T = (\mu,w,p_\mu,p_w) \equiv (\psi_1,\psi_2,\psi_3,\psi_4),\]
where $T$ denotes a transpose operation.
Then, the preceding Equations~(\ref{wHam1})--(\ref{wHam4}) can be compactly expressed as
\begin{equation}
\label{linearBM}
\dot \Psi = {\cal R}\Psi + {\cal I},
\end{equation}
where $\cal R$ is a $4\times 4$ matrix identified as
\begin{equation}\label{relaxR}
{\cal R} =
\left(
\begin{array}{cccc}
-\gamma_\mu    & s             &   1/m_\mu     & 0       \\
s              & -\gamma_w     &   0           & 1/m_w   \\
0              & 0             &  \gamma_\mu   & -s      \\
0              & 0             &   -s          & \gamma_w
\end{array}
\right),
\end{equation}
and the inhomogeneous vector ${\cal I}$ is identified to be
\begin{equation}\label{sourceS}
{\cal I}^T = (\gamma_\mu \mu_d, \gamma_w w_d, 0, 0).
\end{equation}
Equation~(\ref{linearBM}) can be formally integrated to bring about the solution:
\begin{equation}
\label{formalSol}
\Psi(t) = e^{\int_0^t{\cal R}(t^\prime)dt^\prime}\Psi(0) + \int_0^t dt^\prime e^{\int_{t^\prime}^t{\cal R}(\tau)d\tau}{\cal I},
\end{equation}
where the first term on the RHS is a homogeneous solution, given the initial condition $\Psi(0)$, and the second term is the inhomogeneous solution, driven by the source ${\cal I}$.
The formal solution represents a continuous path in 4-dimensional phase space, which variationally optimizes the FE objective [Equation~(\ref{wAction})].
Note that the trace of ${\cal R}$ vanishes identically, i.e., $Tr({\cal R})=0$; accordingly, the sum of its eigenvalues must equal zero, which we use as a consistency condition in the numerical calculation presented in Section~\ref{Application}.
In addition, when the presynaptic signal is constant or saturates in time, the fixed point $\Psi_{eq}$ can be obtained analytically:
\begin{equation}
\label{eqpts}
\Psi_{eq}^T = (\frac{\mu_d+s_\infty w_d/\gamma_\mu}{1-s_\infty^2/\gamma_\mu\gamma_w},\frac{w_d+s_\infty\mu_d/\gamma_w}{1-s_\infty^2/\gamma_\mu\gamma_w},0,0),
\end{equation}
where we used the notation $s_\infty=s(t\rightarrow\infty)$.

\section{Numerical Illustration}
\label{Application}
To exemplify the workings of the BM conducting synaptic inference, we numerically integrated Equations~(\ref{wHam1})--({\ref{wHam4}).
The results are presented below.
\subsection{Free Parameters}
\begin{table}[ht]
\caption{Parameter values we used to produce the data} 
\centering 
\begin{tabular}{c c c c c c c} 
\hline\hline 
\textbf{} & $m_\mu$ & $m_w$ & $\gamma_\mu$ & $\gamma_w$ & $\mu_d$ & $w_d$ \\ [0.5ex] 
\hline 
Solid  & 5 & 0.5 & 1 & 0.1 & 5  & 5 \\
Dotted & 5 & 0.5 & 1 & 0.1 & 10 & 0 \\
\hline\hline
\end{tabular}
\label{Parameters}
\end{table}

Six free parameters appear in the BM and need to be fixed for numerical purposes, of which values we choose as displayed in Table~\ref{Parameters}.
The neural masses $m_\mu$ and $m_w$ are a measure of inferential precision defined to be the inverse noise strengths [Equation~(\ref{masses})].
The frictional coefficients denoted by $\gamma_\mu$ and $\gamma_w$ appear in the generative functions [Equations~(\ref{learning_NGmu}) and (\ref{learning_NGw})], which we set $\gamma_\mu=10\gamma_w$ to account for the slower weight dynamics compared to the neuronal activity.
Also, the parameters $\mu_d$ and $w_d$ in the inhomogeneous vector [Equation~(\ref{sourceS})] represent the brain's prior belief about the postsynaptic and weight values before the presynaptic input arrives.

\subsection{Static Presynaptic Input}

\begin{figure}[t]
\begin{center}
\includegraphics[width=13cm]{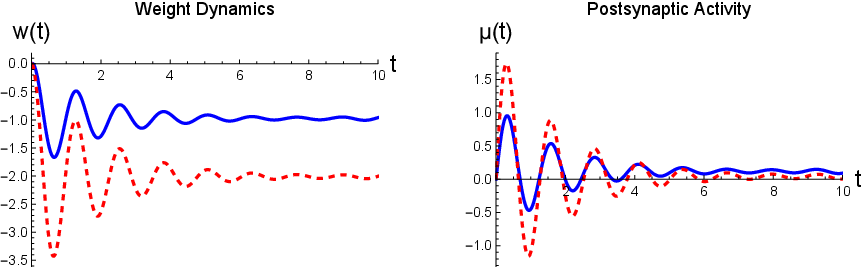}
\caption
{Synaptic dynamics evoked by the static presynaptic input $s=5$. The parameter values that we used to produce the graphs are displayed in Table~\ref{Parameters}; the initial condition was chosen as $\mu(0)=0$ and $w(0)=0$ [All curves are in arbitrary units].
\label{SD_Static}}
\end{center}
\end{figure}

We first present the numerical outcome when the synapse delineated in Figure~\ref{Synapse} is evoked by a static presynaptic signal, which we set as $s=5$.

Figure~\ref{SD_Static} shows the synaptic response of $w$ and $\mu$ to the prescribed input from two different parameter sets displayed in Table~\ref{Parameters}.
Both cases exhibit transient harmonic behaviors: The figure manifests that in response to the static input, the magnitude of the output signals initially increases from the starting value $0$; then, they approach the corresponding fixed points in a sinusoidal manner, $(w_{eq},\mu_{eq})=(-1.0,0.1)$ for the solid curve and $(w_{eq},\mu_{eq})=(-2.0,0.0)$ for the dotted curve.
The transient harmonic behavior is attributed to imaginary eigenvalues of the matrix $R$ for the chosen parameters.
The plots for $p_\mu$ and $p_w$ are not shown because we exploit dynamics near the fixed points, where their values are zero [see Equation~(\ref{eqpts})].
In general, the full synaptic dynamics undergoes in 4-dimensional phase space.

\begin{figure}[t]
\begin{center}
\includegraphics[width=7cm]{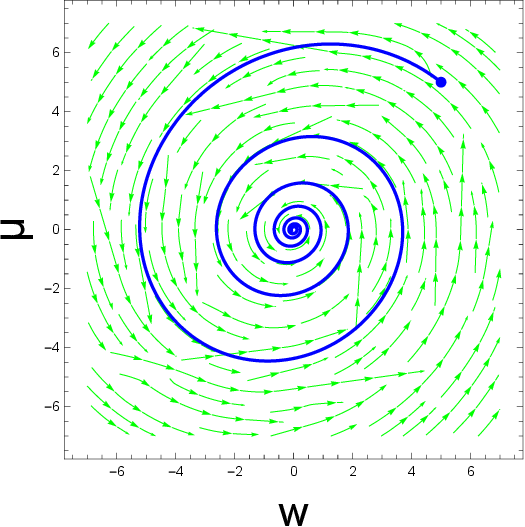}
\caption{
Continuous path driven by the static presynaptic input $s=5$: The initial condition was chosen at $(w,\mu)=(5,5)$, marked by the blue dot; also, for illustrational purposes, we set $\mu_d=0$ and $w_d=0$ while other parameter values are the same as in Table~\ref{Parameters}.
\label{SD_Trajectory}}
\end{center}
\end{figure}

In Figure~\ref{SD_Trajectory}, we illustrate a trajectory in the state space spanned by $(w,\mu)$; the numerical conditions are described in the caption.
One can observe the spiral approach to the fixed point $(0,0)$, starting from the initial condition $(5,5)$ arbitrarily chosen for the illustrational purpose.
Again, the momentum representations are not drawn because their values remain near the equilibrium point in the considered linear dynamics.
Note that the irregularity in the background streamlines is due to the noise in the presynaptic input, reflecting the fact that the brain deterministically predicts cognitive outcomes only on average.
The trajectory we have illustrated is critical to understanding the unconscious cognition of weight change and postsynaptic output.
The temporal course is conditioned on the presynaptic input in the Bayesian brain, a crucial context for our research.

\subsection{Nonstationary Presynaptic Inputs}

Here, we present the numerical results of when the nonstationary presynaptic inputs drive the BM.

First, in Figure~\ref{SD_Time1}, we illustrate the weight dynamics and the postsynaptic activity resulting from the sinusoidally varying presynaptic input.
In this case, the continual harmonic driving causes the output signals to retain their oscillatory behavior and not tend to a fixed point.
The output signals exhibit both positive and negative portions because we considered the voltage-dependent plasticity, aiming at the continuous change, which could induce the negative voltage response \cite{Grag:2022}.
In contrast, only positive signals would be produced if we considered the spike-timing-dependent plasticity.
The momentum variables are not drawn because we follow dynamics near the fixed point in neural phase space, where they remain nearly zero.
Also, it needs to be understood that the weight dynamics is ten times slower than the postsynaptic activity because we assumed the postsynaptic signal decays ten times faster (see Table~\ref{Parameters}).
The same interpretation applies to Figure~\ref{SD_Static}.
\begin{figure}[t]
\begin{center}
\includegraphics[width=13cm]{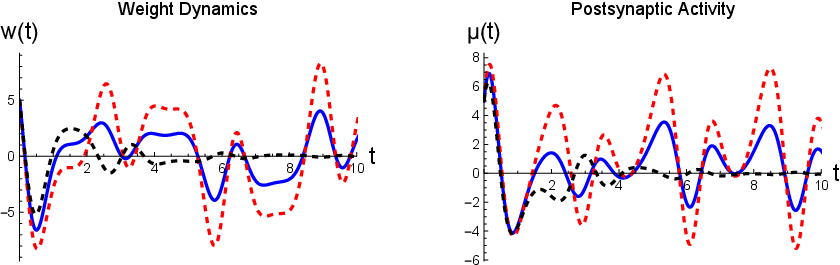}
\caption
{Synaptic dynamics evoked by $s(t)=5\cos t+\eta$, where $\eta$ represents a random fluctuation: The blue solid and red dotted curves are the results from the parameter values displayed in Table~\ref{Parameters}; in addition, we include the black dotted curve from $\mu_d=0$ and $w_d=0$, while other parameter values remain the same. For all data, the initial condition was chosen at $(w,\mu)=(5,5)$ [All curves are in arbitrary units].
\label{SD_Time1}}
\end{center}
\end{figure}

Next, in Figure~\ref{SD_Time2}, we illustrate the neural trajectory in two-dimensional state space produced by the transient input signal that is shown as an inset.
It discloses that the brain's synaptic computation follows a continuous approach to the origin, the fixed point in this case, starting from the chosen initial state $(w,\mu)=(5,5)$.
In numerically integrating the synaptic BM to obtain the trajectory, the parameter values were chosen from Table~\ref{Parameters} except that, for illustrational purposes, the values for $\mu_d$ and $w_d$ were set to be both $0$.
Notice that we did not draw the streamlines in the figure because the presynaptic input is time-dependent, so the streamlines vary at every moment in the trajectory's course.
In Figure~\ref{SD_Trajectory}, by contrast, the input was static so that we could delineate the streamlines.
Notably, our BM theory allowed us to handle the nonautonomous problem induced by the nonstationary presynaptic inputs.
On the other hand, the computability of the usual minimization scheme for the present problem is questionable because the FE landscape is non-static, not allowing a GD implementation, as described in Section~\ref{Hypothesis3}.

\begin{figure}[t]
\begin{center}
\includegraphics[width=7cm]{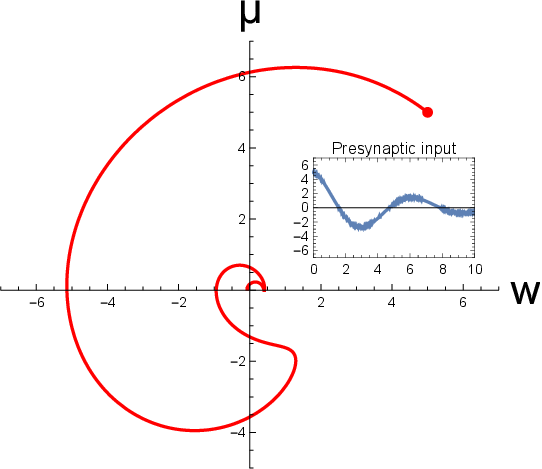}
\caption{
Continuous trajectory in neural state space. The inset shows the transient input signal driving synaptic dynamics, $s(t)=5 e^{-t/5}\cos t +\eta$, where $\eta$ denotes a noise. The initial values of the weight $w$ and postsynaptic signal $\mu$ were chosen at $(w,\mu)=(5,5)$, marked by a red dot; the neural trajectory manifests a continuous approach to the fixed point $(0,0)$ [All curves are in arbitrary units].
\label{SD_Time2}}
\end{center}
\end{figure}

The neural paths we numerically illustrated in the current Section are an optimal trajectory minimizing synaptic surprisal; the actual minimization was performed on the variational FE objective [see Equation~(\ref{wAction})] under the FEP [see Equation~(\ref{FEP2})].
Here, we emphasize that the learned trajectories were self-organized, given the values of material parameters of the brain matter.
In essence, the brain's learning process is unsupervised in the language of machine learning; this means that the brain does not require any external \textit{label} to guide its learning, demonstrating its self-learning efficiency.

\section{Discussion}
\label{Discussion}
The idea that the brain is a neural observer (reinforcer or problem-solving matter) is implicit in the brain-inspired FEP, capable of perceiving, learning, and acting on the external and internal milieus; this renders the FEP a purposive theory.
On the other hand, brain functions must emerge from the brain matter obeying physical laws and principles; the neural substrates afford the biological base for the brain's high-order capability.
Thus, it is significant to recognize that the working FE objective is not a single measure but an architecture hybridizing the teleological rationale and biophysical realities.

In this paper, we continued our endeavor on the continuous-state implementation of the promising FEP as a universal biological principle.
Specifically, we applied our theory to synaptic learning and exemplified the learning dynamics as an inference problem under the FEP.
The noteworthy contributions from our effort are discussed below:

(i) Equation~(\ref{wAction}) is the FE objective in our theory, which suggests that the FE conducts itself as a classical action in Hamilton's principle, i.e., ${\cal S}=F$. We obtained the result by deriving the Onsager-Machlup representations for the NEQ generative densities and inserting them into the Laplace-encoded FE. In our previous studies \cite{Kim:2018,Kim:2021,Kim:2023}, by contrast, the action was identified as a time-integral of the FE, ${\cal S}=\int Fdt$, under the ergodic assumption. The ergodicity asserts that the ensemble average of surprisal over the sensory evidence equals the corresponding temporal average; however, it is difficult to justify the ergodicity idea in the brain.
In the present work, we bypassed the ergodic assumption using the more physics-grounded NEQ densities and avoided employing the generalized coordinates of motion; this makes the FEP computability we proposed more physics-grounded than the other conventional approach.

(ii) The weight variables change in time due to biophysical factors such as an opening of channels at the synapse, through which neurotransmitters transfer in a complex time-dependent manner. Accordingly, we treated the synaptic weights $w$ as a dynamical variable co-evolving with the state variables in completing the synaptic BM. In contrast, the weights are handled as a static parameter in the widely exercised ANNs in machine learning. Also, in the frameworks of ANNs, a nonlinear activation scheme, e.g., the sigmoidal function or ReLU (rectified linear unit), rectifies the network output value \cite{LeCun:2015}. Our biophysics-informed treatment does not use engineering manipulation to regulate the outcome; instead, the learning smoothly follows the continuous BM. We add that one may employ different biophysical models from our Langevin dynamics, such as Izhikevich neurons \cite{Izhikevich} at the neuronal level or neural field models on a mesoscopic scale \cite{Deco:2008}, and apply our framework to derive a desired BM.

(iii) The momentum representations we unveiled [see Equations~(\ref{prerrormu}) and (\ref{prerrorw})] match with the theoretical construct of prediction error in predictive coding theory \cite{Rao:1999,Shipp:2016}. Recently, empirical evidence of error neurons was reported, which encodes prediction errors in the mouse auditory cortex \cite{Audette:2023}. Such a finding provides a neural base for our theory. However, the differentiation between the predictive and error units within a cortical column is still controversial, mainly because of insufficient electrophysiological recordings. Although there is no concrete agreement, the compartmental neuron hypothesis seems to suit the neuronal scenario of functional distinguishability \cite{Larkum:2013,Gillon:2024}, which argues that pyramidal neurons in the cortex are functionally organized such that feedback and feed-forward signals are sent to outer layers (L1) and middle layers (L5), respectively. In this case, our state representations correspond to feedback channels via apical dendrites and momentum representations to feed-forward channels via basal dendrites in somas. The Hebbian sign of Equation~(\ref{learning_NGw}) can be either positive or negative; one can implement spiking predictive coding with the former and the dendrite predictive coding with the latter.

(iv) Data learning via ANNs has become a formidable scientific tool \cite{Schmidhuber:2015}, and much attention is drawn to theoretical questions on how and why they work \cite{Zdeborova:2020}.
This paper suggested that the brain-inspired FEP underlies the widely used $L_2$ objective in machine learning algorithms. The $L_2$ minimization is implemented using a GD with respect to the weights connecting layer by layer, rendering back-propagation of the input-output error reduction in a feed-forward architecture.
However, strictly speaking, the validity of GD updating is limited to situations when the inputs are static or quasi-static. For continuous nonstationary inputs such as a video stream, a bidirectional, recurrent NN (RNN) is employed \cite{Sherstinsky:2020}; RNN sends converted time-series inputs to a pre-structured deep network and performs a GD by incorporating a feedback loop to predict the sequential outputs.
Our BM formulation, in contrast, handles nonstationary learning in a genuinely continuous manner, offering a fresh perspective. The brain integrates the BM to learn a continuous optimal trajectory in neural phase space by minimizing the FE objective rather than estimating a sequential output.
We hope that our physics-guided approach will provide further useful insight into the practice of ANN methodologies in continuous time.

\section{Conclusion}
\label{Conclusion}
Within the continuous-state FEP framework, we cast synaptic learning into minimizing the FE objective, the upper bound for synaptic surprisal in the brain, and derived the BM implementing the FE minimization in a physics-guided manner.
Consequently, we revealed that the brain conducts synaptic learning by integrating the BM to find an optimal trajectory in the reduced-dimensional neural phase space.


\end{document}